\input harvmac
\input epsf

\def\mpla{{Mod.\ Phys.\ Lett.\ }{\bf A}}
\def\npb{{Nucl.\ Phys.\ }{\bf B}}

\def\plb{{Phys.\ Lett.\ }{\bf B}}
\def\prd{{Phys.\ Rev.\ }{\bf D}}

\def\ptp{Prog.\ Th.\ Phys.\ }

\def\half{{\textstyle{1\over2}}}
\def\frak#1#2{{\textstyle{{#1}\over{#2}}}}  
\def\frakk#1#2{{{#1}\over{#2}}}
\def\taub{{\overline{\tau}}}
\def\bbar{{\overline{b}}}
\def\tbar{{\overline{t}}}
\def\Atbar{{\overline{A}_t}}
\def\Abbar{{\overline{A}_b}}
\def\Ataubar{{\overline{A}_{\tau}}}
\def\sy{supersymmetry}
\def\sic{supersymmetric}

\def\lf{16\pi^2}
\def\llf{(16\pi^2)^2}

\def \in{\leftskip = 40 pt\rightskip = 40pt}
\def \out{\leftskip = 0 pt\rightskip = 0pt}
\def\tf{\tilde f}
\def\rt{\tilde r}
\def\tDelta{\tilde\Delta}
\def\tY{\tilde Y}

{\nopagenumbers
\line{\hfil LTH458}
\line{\hfil hep-ph/9909570}
\line{\hfil Revised Version}
\vskip .5in    
\centerline{\titlefont Quasi-infra-red fixed points 
and} 
\centerline{\titlefont renormalisation group invariant trajectories for} 
\centerline{\titlefont non-holomorphic  soft supersymmetry breaking}

\vskip 1in
\centerline{\bf I.~Jack and D.R.T.~Jones}
\medskip
\centerline{\it Dept. of Mathematical Sciences,
University of Liverpool, Liverpool L69 3BX, UK}
\vskip .3in

In the MSSM the quasi-infra-red fixed point for the top-quark  Yukawa
coupling gives rise to specific predictions for the soft-breaking 
parameters. We discuss the extent to which these predictions are
modified  by the introduction of additional ``non-holomorphic''
soft-breaking terms. We also show that in a specific class of theories,
there exists an
RG-invariant trajectory for the ``non-holomorphic'' terms,
which can be understood using a holomorphic spurion term.

\Date{Sept 1999}}

\newsec{Introduction}

The enduring popularity of the MSSM derives originally from the 
demonstration that it  gave rise to gauge coupling unification, 
at a scale consistent with proton decay limits (at least with regard 
to contributions from dimension 6 operators). This success is predicated on  
(or at least consistent with) the desert hypothesis, whereby 
the next fundamental physics scale beyond the weak scale is far beyond it:
gauge unification, a string scale, or even the Planck mass. 
Within this context, a ``standard'' picture of the origin of \sy\ breaking 
has emerged: \sy\ is broken (dynamically or spontaneously) in a distinct 
sector of the theory and transmitted to observable physics via a 
``messenger sector''. At energies below 
a characteristic mass scale $M$ 
the observable effective field theory can be expanded in powers of 
$1/M$; then we suppose that the breaking of \sy\ can be parametrised 
by the vacuum expectation value of the $F$-term of a chiral superfield 
${\cal Z}$, such that $< F_{\cal Z} >\approx M_Z M$, 
and it is easy to show that the following     
soft terms are $O(M_Z)$:
\eqn\Aafo{
L^{(1)}_{\rm SOFT}=(m^2)^j{}_i\phi^{i}\phi_j+
\left(\frak{1}{6}h^{ijk}\phi_i\phi_j\phi_k+\half b^{ij}\phi_i\phi_j
+ \half M\lambda\lambda+{\rm h.c.}\right)}
whereas the following further possible dimension 3 terms are suppressed by 
powers of $M_Z/M$:
\eqn\Aaf{
L^{(2)}_{\rm SOFT}=\half r_{i}^{jk}\phi^i\phi_j\phi_k
+\half m_F{}^{ij}\psi_i\psi_j
+ m_A{}^{ia}\psi_i\lambda_a
+{\rm h.c.}}

The terms in Eq.~\Aaf\ arise from non-holomorphic terms ($D$-terms) in
the effective  field theory, so we will refer to them as non-holomorphic
soft terms  (an abuse of terminology, in fact, inasmuch as of course 
the first 
term in Eq.~\Aafo\ also arises from a non-holomorphic term). 

In fact, if there are no gauge singlets,  the terms in Eq.~\Aaf\ are
``natural'' in the same sense as those of Eq.~\Aafo, in that they do not
give rise to  quadratic divergences; but in any event (within the
paradigm described  above) one would not exclude them even if they do
give quadratic divergences, since we only require naturalness up to the
scale $M$. This was  emphasised recently by
Martin\ref\steve{S.P.~Martin, hep-ph/9907550},  who also pointed out that
by  the same token there are dimension-$4$ \sy-breaking 
contributions which (although suppressed by more powers of $1/M$) may give
rise to  interesting effects.

Returning to the terms shown in Eq.~\Aaf, however, there are two reasons 
why we  should consider them. Firstly, their suppression compared 
to Eq.~\Aafo\ is founded on a specific framework for the 
origin of \sy\ breaking which may or may not be true; secondly, even 
given the framework, the recent model-building trend has been away 
from the desert hypothesis: for example, in the suggestion 
of (very) large extra dimensions. It is not clear to us whether 
in such theories the suppression of Eq.~\Aaf\ relative to Eq.~\Aafo\ will 
necessarily be sustained. Be that as it may, we believe that there is a 
case for an agnostic approach to \sy-breaking whereby 
all dimension 2 and dimension 3 terms are considered without prejudice,
in theories where they do not cause quadratic divergences.  

In a previous paper\ref\jjnss{I.~Jack and D.R.T.~Jones, \plb 457 (1999)
101}\ we gave the one-loop $\beta$-functions for the  parameters defined
in Eq.~\Aaf, both in general and in the  MSSM context. In this paper we
extend the general results to two loops.  We  find (and verify through 
two loops) a RG-invariant relation which can be imposed
between $r$, $b$, $m^2$ and  $m_A$. 
We also investigate the consequences of 
Yukawa infra-red (and quasi-infra-red) fixed point structure  
for the MSSM, 
where we find that some (but not  all) of the predictions founded
on the MSSM survive in the presence  of the non-holomorphic terms. 

\newsec{The $\beta$-functions}

We begin with the one-loop $\beta$-functions for a theory with 
\eqn\Ac{
L=L_{\rm SUSY}+L_{\rm SOFT},}
where
\eqn\Lsoft{
L_{\rm SOFT} = L^{(1)}_{\rm SOFT} + L^{(2)}_{\rm SOFT},}
and where $L_{\rm SUSY}$ is the Lagrangian for  the supersymmetric
gauge theory, containing the gauge multiplet $\{A_{\mu},\lambda\}$ ($\lambda$
being the gaugino) and a chiral superfield $\Phi_i$ with component fields
$\{\phi_i,\psi_i\}$ transforming as a (in general reducible)
representation $R$ of the gauge group $\cal G$. (We give results here for a 
simple gauge group, though the extension to a non-simple gauge group is
straightforward.)
We assume a superpotential of the form
\eqn\Aae{
W=\frak{1}{6}Y^{ijk}\phi_i\phi_j\phi_k.}
Note that we do not include an explicit supersymmetric $\mu$-term in $W$; the 
usual theory containing only $L^{(1)}_{\rm SOFT}$ together with a 
supersymmetric $\mu$-term can be recovered by taking in $L^{(2)}_{\rm SOFT}$
\eqn\muterm{
m_A^{ia}=0, \quad m_F = \mu, \quad r_i^{jk} = Y^{jkl}\mu_{il}} 
and replacing $(m^2)^i{}_j$ in $L^{(1)}_{\rm SOFT}$ by 
$(m^2)^i{}_j + \mu^{il}\mu_{jl}.$
   
The one-loop results for the gauge coupling $\beta$-function $\beta_g$
and  for the chiral field anomalous dimension $\gamma$ are:

\eqn\Aab{
\lf\beta_g =g^3Q \quad\hbox{and}\quad
\lf\gamma^{i}{}_j=P^i{}_j,}
where
\eqn\Aac{
Q=T(R)-3C(G),\quad\hbox{and}\quad
P^i{}_j=\half Y^{ikl}Y_{jkl}-2g^2C(R)^i{}_j.}
Here
\eqn\Aaca{
T(R)\delta_{ab} = \Tr(R_a R_b),
\quad C(G)\delta_{ab} = f_{acd}f_{bcd} \quad\hbox{and}\quad
C(R)^i{}_j = (R_a R_a)^i{}_j,}
and as usual $Y_{ijk}^* = Y^{ijk}$ etc.
For the new soft terms from Eq.~\Aaf\ we have\jjnss:
\eqna\mfij$$\eqalignno{
\lf\beta_{m_{Fij}} &= P^k{}_i m_{Fkj} + P^k{}_j m_{Fik},  &\mfij a\cr
\lf\beta_{m_{Aia}} &= P^j{}_i m_{Aja} + g^2Q m_{Aia}, &\mfij b\cr}$$
and
\eqn\Ac{\eqalign{
\lf(\beta_r)^{jk}_i &=  \half P^l{}_i r^{jk}_l + P^k{}_l r^{jl}_i
+\half r^{mn}_i Y_{lmn}Y^{ljk}+2r^{mj}_l Y_{imn}Y^{kln}
+ 2g^2 r^{jk}_l C(R)^l{}_i\cr &+2g^2r^{mj}_l (R_a)^k{}_i (R_a)^l{}_m
-2m_{Flm}Y^{mnj}Y^{plk}Y_{npi} -4g^2m_{Fil}C(R)^l{}_mY^{mjk}\cr
&- 4g\sqrt{2} \left[g^2C(G) m_A^{ja}(R_a)^k{}_i
+ (R_a)^j{}_l Y^{lmk}Y_{mni}m_A^{na}\right] \quad +(k\leftrightarrow j).\cr}}
For the original soft terms in Eq.~\Aafo\ we have
\eqna\Acc$$\eqalignno{
\lf\beta_h^{ijk}&=U^{ijk}+U^{kij}+U^{jki}, &\Acc a\cr
\lf\beta_b^{ij}&=V^{ij}+V^{ji}, &\Acc b\cr
\lf[\beta_{m^2}]^i{}_j&=W^i{}_j, &\Acc c\cr
\lf\beta_M&=2g^2QM, &\Acc d\cr
}$$
where
\eqna\Aoldm$$\eqalignno{
U^{ijk}&=h^{ijl}P^k{}_l+Y^{ijl}X^k{}_l, &\Aoldm a\cr
V^{ij}&=b^{il}P^j{}_l+ r^i_{lm}h^{jlm}
+r^{im}_l r^{jl}_m -m_{Fkl}Y^{ilm}m_{Fmn}Y^{jnk} \cr &
+4g^2Mm_F^{ik}C(R)^j{}_k -4g^2C(G)m_A^{ia}m_A^{ja}, &\Aoldm b \cr
W^i{}_j&= \half Y_{jpq}Y^{pqn}(m^2)^i{}_n
+\half Y^{ipq}Y_{pqn}(m^2)^n{}_j
+2Y^{ipq}Y_{jpr}(m^2)^r{}_q
+h_{jpq}h^{ipq}\cr &
+r^{kl}_{j}r_{kl}^i +2r^k_{jl}r^{il}_k
-4(m_F^{kl}m_{Flm}+ m_{Ama}m_A^{ka})Y^{imn}Y_{jkn}\cr
&-8g^2(MM^*C(R)^i{}_j + m_F^{kl}m_{Fjk}C(R)^i{}_l +C(G)m_A^{ia}m_{Aja}
+(R_a R_b)^i{}_j m_{Aka}m_A^{kb})\cr
&-4\sqrt{2}g(Y^{iml}m_{Fmn}(R_a)^n{}_j m_{Ala}
+Y_{jml}m_F^{mn}(R_a)^i{}_n m_A^{la})&\Aoldm c \cr}$$
with
\eqn\Aab{
X^i{}_j=h^{ikl}Y_{jkl}+4g^2MC(R)^i{}_j.}
Note that we have omitted from Eq.~\Acc{c}\  a contribution of the form
$g^2(R_a)^i{}_j\Tr[R_am^2]$. This term arises only for $U(1)$ and
amounts to a renormalisation of   the linear $D$-term that is allowed in
that case. The two-loop $\beta$-functions are listed in the Appendix 
(for the case $m_F = 0$). 

There has been much interest recently in RG-invariant relations expressing
the usual soft couplings $M$, $h^{ijk}$ and $(m^2)^i{}_j$ in terms of the
$\beta$-functions for the unbroken theory. 
In Refs.~\ref\con{L. Randall and R. Sundrum, hep-th/9810155\semi
G.F. Giudice, M.A. Luty, H. Murayama and  R. Rattazzi,
JHEP 9812 (1998) 027\semi 
A. Pomarol and  R. Rattazzi, JHEP 9905 (1999) 013\semi
T. Gherghetta, G.F. Giudice, J.D. Wells, hep-ph/9904378\semi
Z. Chacko, M.A. Luty, I. Maksymyk and E. Ponton, hep-ph/9905390\semi
E. Katz, Y. Shadmi and Y. Shirman, JHEP 9908 (1999) 015} these relations were 
derived from the superconformal anomaly, while in 
Ref.~\ref\jjnew{I.~Jack and D.R.T.~Jones, \plb465 (1999) 148}\ they were derived
using exact results for the soft-breaking $\beta$-functions obtained 
using the spurion formalism. From the latter point of view, there would seem no
{\it a priori} reason to expect such RG-invariant results for the new 
non-standard couplings. The reason for this is that the spurion formalism
enables us to relate the renormalisation of the standard soft 
terms $M$, $h^{ijk}$ and $(m^2)^i{}_j$ to the anomalous dimension  $\gamma$ 
of the chiral superfield. This does not carry over to, for example, the 
case of $r_i^{jk}$ because the corresponding 
superspace interaction is $\Phi^2\Phi^*$ which is nonrenormalisable 
and hence leads to divergences beyond those described by $\gamma$. 
It is (at first sight) surprising, therefore, that it is in fact possible to 
develop RG-invariant expressions for the non-standard
couplings. We start by writing $m_F=\mu$
in Eqs.~\mfij{}--\Aoldm{}, since, as
we shall explain in more detail later, $m_F$ will effectively be playing the 
r\^ole of a supersymmetric $\mu$-term. Then firstly, the relation 
\eqn\rfin{r_i^{jk}  = \sqrt{2}g\left[(R_a)^j{}_i m_A^{ka}
+ (R_a)^k{}_i m_A^{ja}\right]+Y^{jkl}\mu_{il}}
defines a renormalisation-group trajectory for $r^{jk}_i$.
If we impose Eq.~\rfin\ in Eq.~\Ac,
we find 
\eqn\RGa{\eqalign{
(\beta_r)^{jk}_i &= \sqrt2\beta_g\left[(R_a)^j{}_i m_A^{ka}
+ (R_a)^k{}_i m_A^{ja}\right]+\sqrt2g\left[(R_a)^j{}_i \beta_{m_A}^{ka}
+ (R_a)^k{}_i \beta_{m_A}^{ja}\right]\cr
&+\beta_Y^{jkl}\mu_{il}+Y^{jkl}\beta_{\mu il}.\cr}}
This clearly implies that Eq.~\rfin\ is RG-invariant. 
Now suppose that in the usual theory, with a supersymmetric $\mu$ term and 
only the soft terms contained in $L^{(1)}_{\rm SOFT}$, we have solved the 
RG equations, with the functions
$(m^2_s)^i{}_j$ and $b_s^{ij}$ being the solutions for 
$(m^2)^i{}_j$ and $b^{ij}$. 
If we additionally impose 
\eqn\RGb{
b^{ij}=b_s^{ij}+2m_A^{ai}m_A^{aj},}
we find, on imposing Eq.~\RGb\ in Eq.~\Acc{b},
\eqn\RGc{
\beta_b^{ij}=\mu {d\over{d \mu}}b_s^{ij}+2\beta_{m_A}^{ai}m_A^{aj}
+2m_A^{ai}\beta_{m_A}^{aj},}
which implies that Eq.~\RGb\ is RG-invariant. Finally, if we set
\eqn\RGd{
m^{Aia} m_{Aja} = \rho\delta^i{}_j, \qquad
(m^2)^i{}_j=(m_s^2)^i{}_j+\mu^{ik}\mu_{kj}+2\rho\delta^i{}_j}
where $\rho$ is an arbitrary constant, and the matter multiplet 
satisfies $C(R)^i{}_j = C(G) \delta^i{}_j$,
then we find on substituting Eq.~\RGd\ into Eq.~\Acc{c}\ that 
\eqn\RGe{
(\beta_{m^2})^i{}_j = \mu {d\over{d \mu}}(m_s^2)^i{}_j
+\beta_{\mu}^{ik}\mu_{kj}+\mu^{ik}
\beta_{\mu kj}+2\beta_{m_A}^{ai}m_{Aaj}+2m_A^{ai}\beta_{m_Aaj},}
demonstrating the RG-invariance of Eq.~\RGd. Note that here we are  
including a supersymmetric $\mu$-term. To be more explicit, another way to 
phrase our results is to say that in a 
theory with $W=\frak{1}{6}Y^{ijk}\phi_i\phi_j\phi_k
+\frak12\mu^{ij}\phi_i\phi_j$, together with $L_{\rm SOFT}$ as in
Eq.~\Lsoft (but taking $m_F = 0$ in Eq.~\Aaf), the relations
\eqna\RGee$$\eqalignno{
r_i^{jk}  &= \sqrt{2}g\left[(R_a)^j{}_i m_A^{ka}
+ (R_a)^k{}_i m_A^{ja}\right], &\RGee a\cr 
b^{ij} &=b_s^{ij}+2m_A^{ai}m_A^{aj},&\RGee b\cr
m^{Aia} m_{Aja} &= \rho\delta^i{}_j, 
\qquad (m^2)^i{}_j = (m_s^2)^i{}_j+2\rho\delta^i{}_j&\RGee c\cr}$$
are RG-invariant (once again with the proviso that the matter multiplet
satisfies $C(R)^i{}_j = C(G) \delta^i{}_j$ in the case of Eq.~\RGee{c}). 
Using the two-loop results 
given in the Appendix, we can show that the trajectory is also RG-invariant at 
two-loop order. In the special case of a one-loop 
finite theory (and setting $\mu = 0$) 
the above trajectory was described in Ref.~\jjnss. 

The existence of the RG trajectory described by Eq.~\RGee{}\ can 
in fact be understood using spurions.
\foot{We are most grateful to the referee 
for indicating to us the following argument.} Consider the term
\eqn\spuri{
L_{\rm SOFT}= \sqrt{2}m_A\int \theta^{\alpha}W^a_{\alpha}\Phi^a\, d^2\theta
\quad +\hbox{c.c.}}
where $\Phi^a (\phi, \psi, F)$ 
is a chiral superfield in the adjoint representation 
and 
\eqn\fldstg{W^a_{\alpha} = \lambda^a_{\alpha} - D^a\theta_{\alpha}+\cdots}
is the usual superspace gauge field strength.
In the Wess-Zumino gauge this reduces to 
\eqn\spuria{
L_{\rm SOFT} = m_A (\lambda^a\phi^a +\hbox{c.c.}) - 
\sqrt{2}m_AD^a (\phi^a + \phi^{*a}).}   
When the auxiliary field $D$ is eliminated this produces 
the following contributions to the Lagrangian:
\eqn\spurib{
L =  m_A (\lambda^a\psi^a +\hbox{c.c.})
+\frak{1}{2}\left[ g\phi^* R^a\phi + \sqrt{2}m_A (\phi^a + \phi^{*a})\right]^2}
which, it is easy to see, precisely accounts for all the terms in 
Eq.~\RGee{}. The fact that we were forced to place the chiral 
superfield in the adjoint representation to obtain an RG 
invariant trajectory is now simply understood 
in that for such a field we can obtain all our ``non-holomorphic'' 
soft breakings from a single {\it holomorphic\/} term, Eq.~\spuri. Moreover, 
the fact that it {\it is\/} holomorphic means that we can immediately 
apply the non-renormalisation theorem to show that (on the trajectory)
\eqn\allorders{
\beta_{m_A} = \frakk{\beta_g}{g} + \gamma.}
It is easy to verify this result through two loops using Eqs.~(A2), \mfij{b}.
\newsec{The MSSM}

Retaining only the third generation Yukawa couplings we have 
the superpotential

\eqn\wmssm{W = \lambda_t H_2
Q \tbar + \lambda_b H_1 Q \bbar  + \lambda_{\tau}H_1 L \taub,}
and soft breaking terms 

\eqn\smssma{\eqalign{ L^{(1)}_{\rm SOFT} &= \sum_{\phi}
m_{\phi}^2\phi^*\phi + \left[m_3^2 H_1
H_2 + \sum_{i=1}^3\half M_i\lambda_i\lambda_i  + {\rm h.c. }\right]\cr 
&+ \left[A_{t}\lambda_t H_2 Q \tbar  +
A_b\lambda_b H_1 Q \bbar  + A_{\tau}\lambda_{\tau}H_1 L \taub
+ {\rm h.c. }\right]\cr}}
and
\eqn\smssmb{ L^{(2)}_{\rm SOFT} = m_{\psi} \psi_{H_1}\psi_{H_2} +
\Atbar\lambda_t H_1^* Q \tbar
+ \Abbar\lambda_b H_2^* Q \bbar
+ \Ataubar\lambda_{\tau}H_2^* L \taub   + {\rm h.c. }}
If we set $m_{\psi} = \Ataubar = \Abbar = \Atbar = \mu$ and 
$m_{1,2}^2\rightarrow m_{1,2}^2+\mu^2$ then we recover the 
MSSM. (A note on notation: in our previous paper\jjnss\ we followed 
Inoue et al.\ref\inetal{K.~Inoue 
et al, \ptp 67 (1982) 1889;  erratum {\it ibid\/} 70 (1983) 330}, who  
used $m_{\psi}=m_4$, $\Ataubar=m_5$, $\Abbar=m_7$, $\Atbar=m_9$, and
correspondingly $A_{\tau}=m_6$, $A_b=m_8$ and $A_t=m_{10}$.) 
As in Eq.~\wmssm\ we assume 3rd. generation dominance here (this may 
not be true, of course). In fact we neglect all mixing between the 
generations  and all couplings associated with the first two generations 
throughout; for the generalisation to include these (in the absence 
of our non-holomorphic terms) in the quasi-fixed-point context, 
see Ref.~\ref\abel{S.A.~Abel and B.C.~Allanach, \plb415 (1997) 371}.  

The supersymmetric couplings evolve according to the well-known equations
\eqna\susyev$$\eqalignno{
{d\alpha_i\over{dt}} &= -b_i\alpha_i^2, \quad (i=1,2,3) &\susyev a \cr
{dy_t\over{dt}} &= -y_t(6y_t+y_b-\sum_iC^t_i\alpha_i),&\susyev b \cr
{dy_b\over{dt}} &= -y_b(6y_b+y_t+y_{\tau}-\sum_iC^b_i\alpha_i),&\susyev c \cr
{dy_{\tau}\over{dt}} &= -y_{\tau}(4y_{\tau}+3y_b-\sum_iC^{\tau}_i\alpha_i),
&\susyev d \cr}$$
where $t=-{1\over{2\pi}}\ln\mu$, 
\eqn\defs{
\alpha_i={g_i^2\over{4\pi}},\qquad y_t={\lambda_t^2\over{4\pi}}
\qquad\hbox{etc.}}
and 
\eqn\defsa{\eqalign{
b_i=\left(\frak{33}{5},1,-3\right), \quad C^t_i &= \left(\frak{13}{15},3,\frak{16}{3}
\right),\cr
C^b_i=\left(\frak{7}{15},3,\frak{16}{3}\right),\quad C^{\tau}_i &= \left(
\frak95,3,0\right),\quad i=1,2,3.\cr}}
It is straightforward to show from our results that
\eqna\fourb$$\eqalignno{
{dm_{\psi}\over{dt}}&= -\frak12(y_{\tau} + 3 y_b
+ 3y_t- 2\sum_iC^H_i\alpha_i)m_{\psi}
, &\fourb a\cr
{d\Ataubar\over{dt}}&= -\frak12(y_{\tau} - 3 y_b + 3y_t)\Ataubar
-3y_b\Abbar+(2m_{\psi}-\Ataubar)\sum_iC^H_i\alpha_i, &\fourb b\cr
{d\Abbar\over{dt}}&= -\frak12(3y_b + 5y_t -y_{\tau})\Abbar
-\Ataubar y_{\tau} +y_t(2m_{\psi}-\Atbar)\cr
&+(2m_{\psi}-\Abbar)\sum_iC^H_i\alpha_i, &\fourb c\cr
{d\Atbar\over{dt}}&= -\frak12(y_{\tau} + 5 y_b + 3y_t)\Atbar
+y_b(2m_{\psi}-\Abbar ) + (2m_{\psi}-\Atbar )
\sum_iC^H_i\alpha_i,&\fourb d\cr
{dA_{\tau}\over{dt}} &= -4y_{\tau}A_{\tau} -3y_b A_b 
-\sum_iC^{\tau}_i\alpha_iM_i, &\fourb e\cr
{dA_b\over{dt}} &= -y_{\tau}A_{\tau} -6y_b A_b
- y_t A_{t}-\sum_iC^b_i\alpha_iM_i, &\fourb f\cr
{dA_{t}\over{dt}} &= -y_b A_b
- 6y_t A_{t} -\sum_iC^t_i\alpha_iM_i, &\fourb g\cr
{dm_1^2\over{dt}}&= 
-y_{\tau}(m_1^2 + A_{\tau}^2 + m_L^2 + m_{\taub}^2)
-3y_b(m_1^2 + A_b^2 + m_Q^2 + m_{\bbar}^2)\cr
&-3y_t \Atbar^2 
+2\sum_iC^H_i\alpha_i(m_{\psi}^2 +M_i^2), &\fourb h\cr
{dm_2^2\over{dt}}&= 
-3y_t (m_2^2 + A_{t}^2 + m_Q^2 + m_{\tbar}^2)
-y_{\tau}\Ataubar^2-3y_b\Abbar^2\cr
&+2\sum_iC^H_i\alpha_i(m_{\psi}^2 +M_i^2), &\fourb i\cr
{dm_3^2\over{dt}}&=
-\frak12(y_{\tau}+3y_b+3y_t)m_3^2
-y_{\tau}\Ataubar A_{\tau}-3y_b\Abbar A_b-3y_t\Atbar A_t
\cr &+\frak12\sum_iC^H_i\alpha_i(m_3^2 -2M_im_{\psi}), &\fourb j\cr
{dm_Q^2\over{dt}}&= 
-X_b -X_t +2\sum_iC^Q_i\alpha_iM_i^2, &\fourb k\cr
{dm_{\tbar}^2\over{dt}} &= 
-2X_t +2\sum_iC^{\tbar}_i\alpha_iM_i^2, &\fourb l\cr
{dm_{\bbar}^2\over{dt}} &=
-2X_b+2\sum_iC^{\bbar}_i\alpha_iM_i^2, &\fourb m\cr
{dm_{L}^2\over{dt}} &=
-X_{\tau}+2\sum_iC^{H}_i\alpha_iM_i^2, &\fourb n\cr
{dm_{\taub}^2\over{dt}} &=
-2X_{\tau}+2\sum_iC^{\taub}_i\alpha_iM_i^2, &\fourb o\cr
{dM_i\over{dt}} &= -b_iM_i\alpha_i, &\fourb p\cr }$$
where 
\eqn\Cedf{\eqalign{
C^Q &=\left(\frak{1}{30},\frak32,\frak83\right),\quad C^{\tbar}=
\left(\frak{8}{15},0,\frak83\right),
\quad C^{\bbar}=\left(\frak{2}{15},0,\frak83\right),\cr
C^{\taub} &=\left(\frak{6}{5},0,0\right)\quad 
C^H_i=\left(\frak{3}{10},\frak32,0\right), \quad i=1,2,3\cr}} 
and where  
\eqn\xdefs{\eqalign{
X_t &=   
y_t(m_Q^2 +  m_{\tbar}^2 + m_2^2 + \Atbar^2 + A_{t}^2  - 2m_{\psi}^2),\cr
X_b &=  y_b (m_Q^2 +  m_{\bbar}^2+  m_1^2 + \Abbar^2 + A_b^2 - 2m_{\psi}^2),\cr
X_{\tau} &=  
y_{\tau} (m_L^2 + m_{\taub}^2+  m_1^2 
+ \Ataubar^2 + A_{\tau}^2 - 2m_{\psi}^2).\cr}}

\subsec{The small $\tan\beta$ regime}

In the small $\tan \beta$ regime where we take $y_b=y_{\tau}=0$, 
Eqs.~\susyev{a,b}\
are easily solved to give
\eqna\ssmak$$\eqalignno{
\alpha_i(t) &= {\alpha_0\over{1+b_i\alpha_0t}},&\ssmak a\cr
y_t (t)  &= y_0f(t)H_6(t,y_0)&\ssmak b\cr }$$
where
\eqn\ssmaj{
f(t) = \prod_i[1+b_i\alpha_0t]^{C^t_i\over{b_i}},}
and 
\eqn\ssmai{H_6(t,y_0)={1\over{1+6y_0F(t)}},\quad
F(t) = \int_0^{t} f(\tau)\,d\tau}
and where $y_0=y_t(0)$ and we assume a common initial gauge coupling 
$\alpha_i(0)=\alpha_0$ at a unification scale $M_U$. 
We then easily solve Eqs.~\fourb{a-d}\ to give
\eqna\sols$$\eqalignno{
m_{\psi}(t) &= H_6(t,y_0)^{\frak14}\tf(t)m_{\psi}(0),&\sols a \cr
{{\cal A}_t(t)} &= 1+\tf(t)^{-2}\left[{{\cal A}_t (0)}
-1\right],&\sols b \cr
{{\cal A}_b}(t) &= 1+H_6(t,y_0)^{\frak16}\tf(t)^{-2}\left[
{{\cal A}_t (0)}
+{{\cal A}_b (0)}-2\right]\cr
&+\tf(t)^{-2}\left[1-{{\cal A}_t (0)}\right], &\sols c \cr
{{\cal A}_{\tau}}(t) &= 1+\tf(t)^{-2}\left[{{\cal A}_{\tau} (0)}-1\right],
&\sols d\cr}$$
where 
\eqn\ftil{
\tf(t)=\prod_i[1+b_i\alpha_0t]^{{C^H_i\over{b_i}}},}
and 
\eqn\caldef{{\cal A}_{t} ={\Atbar(t)\over{m_{\psi}(t)}},\quad 
{\cal A}_{b} ={\Abbar(t)\over{m_{\psi}(t)}},\quad
{\cal A}_{\tau} ={\Ataubar(t)\over{m_{\psi}(t)}}.}
Using the elementary solution of Eq.~\fourb{p},  
\eqn\Msol{
M_i={M_0\over{1+b_i\alpha_0t}},}
where we assume a common initial gaugino mass $M_i(0)=M_0$, 
we can also solve 
Eq.~\fourb{g}, giving  
\eqn\Asol{
A_{t}(t)=\{A_{t}(0)+6y_0M_0[tf(t)-F(t)]\}H_6(t,y_0)-M_0t{1\over{f(t)}}
{df\over{dt}}.}
It is instructive to note that the boundary condition on the gaugino masses
plays a crucial r\^ole in determining the form of the solution. 
Thus if we take instead 
\eqn\conan{M_i(0)=m_{\frak{3}{2}}b_i\alpha_0,}
then we obtain 
\eqn\conana{A_t = H_6\left[A_t (0) + m_{\frak{3}{2}}(6y_0
- \sum_i C_i^t\alpha_0)\right] + 
m_{\frak{3}{2}}\left[\sum_i C_i^t\alpha_i (t) - 6y(t)\right],} 
which, if we impose the initial condition $A_t (0) + m_{\frak{3}{2}}(6y_0  
- \sum_i C_i^t\alpha_0)=0$, is the one-loop form of the conformal anomaly 
solution\con\jjnew\ for $A_t$. 

Proceeding with Eq.~\Asol, we can (with more labour) 
solve Eqs~\fourb{h,i,k-o}, giving
\eqn\msoft{\eqalign{
m_Q^2(t) &= m_Q^2(0)
+M_0^2(\frak83f_3(t)+\frak32f_2(t)+\frak{1}{30}f_1(t))
+\frak16\Delta(t)+Y(t),\cr
m_{\tbar}^2(t) &= m_{\tbar}^2(0)+M_0^2(\frak83f_3(t)+\frak{8}{15}f_1(t))
+\frak13\Delta(t)+2Y(t),\cr
m_{\bbar}^2(t) &= m_{\bbar}^2(0)+M_0^2(\frak83f_3(t)+\frak{2}{15}f_1(t)),\cr
m_2^2(t) &= m_2^2(0)+M_0^2(\frak32f_2(t)+\frak{3}{10}f_1(t))
+\frak12\Delta(t)-3Y(t),\cr
m_1^2(t) &= m_1^2(0)+M_0^2(\frak32f_2(t)+\frak{3}{10}f_1(t))
\cr
&+\{\tf(t)^2m_{\psi}(0)^2+2m_{\psi}(0)[\Atbar(0)-m_{\psi}(0)]\}
H_6(t,y_0)^{\frak12}\cr
&+m_{\psi}(0)[m_{\psi}(0)-2\Atbar(0)]-3y_0[\Atbar(0)-m_{\psi}(0)]^2
\Omega_{(6,\frak32)}(t),\cr
m_L^2(t) &= m_L^2(0)+M_0^2\left(\frak32f_2(t)+\frak{3}{10}f_1(t)\right),\cr
m_{\taub}^2(t) &= m_{\taub}^2(0)+\frak65M_0^2f_1(t)\cr}}
where
\eqn\Deldef{\eqalign{
f_i(t) &= {1\over{b_i}}\left(1-{1\over{(1+b_i\alpha_0t)^2}}\right),\cr
\Delta(t) &= [\Sigma(0)-A_{t}(0)^2]H_6(t,y_0)
+\{A_{t}(0)+6M_0y_0[tf(t)-F(t)]\}^2H_6(t,y_0)^2\cr
&-6y_0M_0^2H_6(t,y_0)t^2{df\over{dt}}-\Sigma(0)\cr
&+\{m_{\psi}(0)^2\tf(t)^2-m_{\psi}(0)[m_{\psi}(0)-\Atbar(0)]\}
H_6(t,y_0)^{\frak12}\cr
&+\{m_{\psi}(0)[2\Atbar(0)-3m_{\psi}(0)]-3[\Atbar(0)
-m_{\psi}(0)]^2\Omega_{(6,\frak12)}(t)\}H_6(t,y_0),\cr}}
with $\Sigma=m_Q^2+m_{\tbar}^2+m_2^2$, and where
\eqn\Omdefs{
\Omega_{(a,n)}(t)=\int_0^tf(\tau)\tf(\tau)^{-2}H_a(\tau,y_0)^nd\tau}
and
\eqn\Ydef{\eqalign{
Y(t) &= -\frak16\{m_{\psi}(0)^2\tf^2+2m_{\psi}(0)[m_{\psi}(0)-\Atbar(0)]\}
H_6(t,y_0)^{\frak12}\cr
&+\frak16m_{\psi}(0)[3m_{\psi}(0)-2\Atbar(0)]
-\frak12y_0[\Atbar(0)-m_{\psi}(0)]^2\Omega_{(6,\frak32)}(t),\cr}}

Once again, use of the alternative boundary condition Eq.~\conan\ and 
the corresponding solution for $A_t(t)$ leads instead (with appropriate initial
conditions for the masses) to the conformal 
anomaly form for the $m^2$ terms. This we leave as an exercise for the 
reader. 

In the special case of the MSSM, 
explicit solutions for the soft parameters were 
written down in 
Refs.~\ref\kom{H. Komatsu, INS-Rep-469, 1983\semi 
L.E. Ib\'a\~nez and C.~Lopez \npb 233 (1984) 511\semi
L.E. Ib\'a\~nez, C.~Lopez and C. Mu\~noz, \npb 256 (1985) 218}. 
Recently Codoban and Kazakov~\ref\kaz{S.~Codoban and D.I.~Kazakov,
hep-ph/9906256}\ 
have given an elegant derivation 
using the spurion formalism; their  results may be obtained 
by setting $m_{\psi} = \Ataubar = \Abbar = \Atbar = 0$. 
We note that in the more general case considered here it is not possible to
obtain a simple closed form for $m_3^2(t)$. However, this is not a major 
drawback since in typical running analyses, $m_3^2(M_Z)$ is in any case 
derived by minimising the effective potential.

\subsec{The large $\tan\beta$ region}

In the large $\tan\beta$ region, if we make the 
approximation\ref\marcela{M.~Carena, M. Olechowski, S. Pokorski
and C.E.M. Wagner,  
\npb426 (1994) 269}
$y_b\approx y_t=y$, $y_{\tau}\approx0$,  
the Yukawa coupling is given to a good approximation by
\eqn\yuk{
y(t) =y_0\hat f(t)H_7(t,y_0),}
where 
\eqn\fhatdef{
\hat f(t) = \prod_i[1+b_i\alpha_0t]^{C^{tb}_i\over{b_i}},}
with $C^{tb}=\left(\frak23,3,\frak{16}{3}\right)$,
and
\eqn\Gdef{H_7(t,y_0)={1\over{1+7y_0\hat F(t)}},\quad
\hat F(t) = \int_0^{t} \hat f(\tau)\,d\tau.}
Note that $C^{tb}_{2,3}=C^t_{2,3}=C^b_{2,3}$ while we have chosen to set  
$C^{tb}_1=\frak12(C^t_1+C^b_1)$. (In fact, it makes very little difference 
if we instead use $C^{tb} = C^t$, in which case $f = \hat f$ 
and $F = \hat F$.) 
We can then solve Eqs.~\fourb{a-d}\ to obtain
\eqna\solsb
$$\eqalignno{
m_{\psi}(t) &= H_7(t,y_0)^{\frak37}\tf(t)m_{\psi}(0),&\solsb a \cr
{{\cal A}_t}(t) &= 1+\frak12\tf(t)^{-2}H_7(t,y_0)^{\frak27}
\left[{{\cal A}_t (0)}+{{\cal A}_b (0)}-2\right]\cr
&+\frak12\tf(t)^{-2}\left[{{\cal A}_t (0)}-
{{\cal A}_b (0)}\right],&\solsb b \cr
{{\cal A}_b}(t) &= 1+\frak12\tf(t)^{-2}H_7(t,y_0)^{\frak27}
\left[{{\cal A}_t (0)}+{{\cal A}_b (0)}-2\right]\cr
&-\frak12\tf(t)^{-2}\left[{{\cal A}_t (0)}-
{{\cal A}_b (0)}\right],&\solsb c \cr
{{\cal A}_{\tau}}(t) &= 1+H_7(t,y_0)^{-\frak37}\tf(t)^{-2}
\left[{{\cal A}_{\tau} (0)}+\frak15{{\cal A}_t (0)}
-\frak45{{\cal A}_b (0)}-\frak25\right]\cr
&+\frak{3}{10}H_7(t,y_0)^{\frak27}\tf(t)^{-2}
\left[{{\cal A}_t (0)}+{{\cal A}_b (0)}-2\right]\cr
&-\frak12\tf(t)^{-2}
\left[{{\cal A}_t (0)}-{{\cal A}_b (0)}
\right].&\solsb d\cr}$$
We also find from  Eqs.~\fourb{f,g}\ that 
\eqn\Asolb{\eqalign{
A_{t}(t) &= \{\frak12(A_{t}(0)+A_b(0))+7y_0M_0[t\hat f(t)-\hat F(t)]\}H_7(t,y_0)
-M_0t{1\over{\hat f(t)}}{d\hat f\over{dt}}\cr
&+\{\frak12(A_{t}(0)-A_b(0))+5y_0M_0[tg(t)-G(t)]\}H_5(t,y_0)
-M_0t{1\over{g(t)}}{dg\over{dt}},\cr
A_{b}(t) &= \{\frak12(A_{t}(0)+A_b(0))+7y_0M_0[t\hat f(t)-\hat F(t)]\}H_7(t,y_0)
-M_0t{1\over{\hat f(t)}}{d\hat f\over{dt}}\cr
&-\{\frak12(A_{t}(0)-A_b(0))+5y_0M_0[tg(t)-G(t)]\}H_5(t,y_0) 
+M_0t{1\over{g(t)}}{dg\over{dt}},
\cr}}
where 
\eqn\solsss{
g=[1+b_1\alpha_0t]^{\frak{c_1}{b_1}},}
with $2c_1=C^t_1-C^b_1=\frak15$, and 
\eqn\ftdef{\eqalign{
G &= {1\over{(c_1+b_1)\alpha_0}}\left\{[1+b_1\alpha_0t]^{{c_1\over{b_1}}+1}-1
\right\},\cr
&H_5(t,y_0)={1\over{1+5y_0\hat F(t)}}.\cr}}
With the further assumptions $\Abbar(0)\approx\Atbar(0)$, $A_b(0)\approx 
A_t(0)$,
$m_1^2\approx m_2^2$, $m_{\bbar}^2\approx m_{\tbar}^2$,
and using $g(t)\approx 1$ and $G(t)\approx t$, Eqs.~\solsb{}, \Asolb\ 
simplify to
\eqna\solsc
$$\eqalignno{
m_{\psi}(t) &= H_7(t,y_0)^{\frak37}\tf(t)m_{\psi}(0),&\solsc a \cr
{{\cal A}_t}={{\cal A}_b} &= 1+\tf(t)^{-2}H_7(t,y_0)^{\frak27}
\left[{{\cal A}_t (0)}-1\right],&\solsc b \cr
{{\cal A}_{\tau}} &= 1+H_7(t,y_0)^{-\frak37}\tf(t)^{-2}
\left[{{\cal A}_{\tau} (0)}-\frak35{{\cal A}_t (0)}
-\frak25\right]\cr
&+\frak35H_7(t,y_0)^{\frak27}\tf(t)^{-2}
\left[{{\cal A}_t (0)}-1\right],&\solsc c \cr
A_{t}(t)=A_b(t) &= \{A_{t}(0)+7y_0M_0[t\hat f(t)-\hat F(t)]\}H_7(t,y_0)
-M_0t{1\over{\hat f(t)}}{d\hat f\over{dt}},&\solsc d\cr}$$
and we find that with these assumptions we can obtain the following explicit
solutions for the soft masses: 
\eqn\msofta{\eqalign{
m_Q^2(t) &= m_Q^2(0)+M_0^2(\frak83f_3(t)+\frak32f_2(t)+\frak{1}{30}f_1(t))
+\frak27\tDelta(t)+\tY(t),\cr
m_{\tbar}^2(t) &= m_{\tbar}^2(0)+M_0^2(\frak83f_3(t)+\frak{8}{15}f_1(t))
+\frak27\tDelta(t)+\tY(t),\cr
m_2^2(t) &= m_2^2(0)+M_0^2(\frak32f_2(t)+\frak{3}{10}f_1(t))+\frak37\tDelta(t)-2\tY(t),\cr}}
where
\eqn\tDeldef{\eqalign{
\tDelta &= [\Sigma(0)-A_{t}(0)^2]H_7(t,y_0)
+[A_{t}(0)+7M_0y_0(t\hat f(t)-\hat F(t))]^2H_7(t,y_0)^2\cr
&-7y_0M_0^2H_7(t,y_0)t^2{d\hat f\over{dt}}-\Sigma(0)\cr
&+\Bigl[m_{\psi}(0)^2(\tf(t)^2H_7(t,y_0)^{-\frak17}-1)
-7y_0[\Atbar(0)-m_{\psi}(0)]^2{\hat \Omega}_{(7,\frak67)}(t)\cr 
&+14m_{\psi}(0)(\Atbar(0)-m_{\psi}(0))(H_7(t,y_0)^{\frak17}-1)\Bigr]H_7(t,y_0)
,\cr}}
with
\eqn\tYdef{
\tY(t)=\frak27m_{\psi}(0)^2\{1-H_7(t,y_0)^{\frak67}\tf(t)^2\}.}
${\hat \Omega}$ is defined like $\Omega$ in Eq.~\Omdefs, except that 
$f\to{\hat f}$.

\subsec{Quasi-infra red fixed points and sum rules}

The possibility that the weak-scale values of various parameters in the MSSM are
governed by quasi-infra-red fixed-point (QIRFP) behaviour
\ref\hil{C.T.~Hill, \prd 24 (1981) 691}
 has received a good
deal of attention; see for example
\abel\kaz\marcela\ref\QIRref{
J.~Bjorkman and D.R.T. Jones, \npb 259 (1985) 533\semi
M.~Carena, S.~Pokorski and C.E.M.~Wagner, \npb406 (1993) 59\semi
M.~Carena et al, \npb369 (1992) 33\semi
M.~Carena et al, \npb 419 (1994) 213\semi
W.A.~Bardeen et al, \plb 320 (1994) 110\semi
M.~Carena and C.E.M.~Wagner, \npb 452 (1995) 45\semi
M.~Lanzagorta and G.G.~Ross, \plb 364 (1995) 363}%
\nref\jjpf{P.M.~Ferreira, I.~Jack and D.R.T.~Jones,
\plb 392 (1997) 376}%
\nref\kaza{G.K. Yeghian, M. Jurcisin and D.I. Kazakov, 
\mpla14 (1999) 601\semi
M.~Jurcisin and D.I.~Kazakov, \mpla14 (1999) 671}%
--\ref\abelb{S.A.~Abel and B.C.~Allanach, hep-ph/9909448}.  
In this scenario, the value of the Yukawa coupling at the weak scale 
is close to the value corresponding to having a Landau pole at the 
unification scale. It follows that this value will be obtained for a 
wide range of input Yukawa couplings at $M_U$. 
In the small $\tan\beta$ case, for example, we have from 
Eq.~\ssmak{b}\ that  when $6y_0F(t) >> 1$ then
$y_t\approx f(t)/6F(t)$, independent of $y_0$. 
Moreover, since 
$F(M_Z)\approx 18$ it follows that there is a range of perturbatively 
believable values of $y_0$ such that the QIRFP is approached at $M_Z$. 
(For a discussion of the extent to which this 
scenario is preserved at higher orders, see 
Ref.~\jjpf.) In what 
follows we will investigate whether this behaviour of the Yukawa 
coupling causes QIRFP behaviour for the  soft parameters, 
simply by taking the   limit of large $y_0$, and examining whether the results 
are independent of the initial conditions at $M_Z$. Of course whether the 
range of $y_0$ corresponding to close approach to any resulting QIRFP 
includes perturbatively believable values will 
depend on the details of the solution. 

Thus from Eq.~\Asol\ we see that for small $\tan\beta$ and large $y_0$,  
\eqn\AQIR{
A_t(t)\approx M_0\left({tf(t)\over{F(t)}}-1-{t\over{f(t)}}{df\over{dt}}\right).}
In the large $\tan\beta$ case, we have from Eq.~\yuk\ that for
large $\tan\beta$, $y\approx \hat f(t)/7\hat F(t)$, 
and from Eq.~\Asolb
\eqn\AQIRa{\eqalign{
A_t\approx &M_0\left({t\hat f(t)\over{\hat F(t)}}-{t\over{\hat f(t)}}
{d\hat f\over{dt}}+ {tg(t)\over{G(t)}}-{t\over{g(t)}}{dg\over{dt}}-2\right) \cr
A_b\approx &M_0\left({t\hat f(t)\over{\hat F(t)}}-{t\over{\hat f(t)}}
{d\hat f\over{dt}}-{tg(t)\over{G(t)}}+{t\over{g(t)}}{dg\over{dt}}\right) \cr
}}
Since the only difference between $f$ and $\hat f$, and correspondingly 
$F$ and $\hat F$, is the replacement of $C^t_1$ by $C^{tb}_1$, and since
we have $g(t)\approx 1$ and $G(t)\approx t$, we see that the QIRFP 
predictions for $A_t$
and $A_b$ for large $\tan\beta$ are in fact close to the small $\tan\beta$
prediction for $A_t$. To be more explicit, for small $\tan\beta$ we find
\eqn\QIRAnum{
{A_t(M_Z)\over{M_3(M_Z)}}\approx-0.6,}
with less than a $1\%$ difference in the large $\tan\beta$ case for 
${A_t(M_Z)\over{M_3(M_Z)}}$ or ${A_b(M_Z)\over{M_3(M_Z)}}$, in agreement with
Refs.~\abel\marcela\kaza. 

Turning to the soft masses, 
we find that for small $\tan\beta$ and large $y_0$ 
\eqn\QIRa{
\Delta\approx M_0^2\left(
{(tf-F)^2\over{F^2}}-{t^2\over F}{df\over {dt}}\right)
-\Sigma(0)}
with a similar equation for $\tDelta$ in the large $\tan\beta$ case,
but with $f\rightarrow\hat f$, $F\rightarrow \hat F$
(after setting $A_b\approx A_t$, $\Abbar\approx\Atbar$), so that $\Delta$
depends on the initial values of the soft masses through $\Sigma(0)$.

In the standard case where the superpotential Eq.~\wmssm\ contains also a $\mu$ 
term, but the soft terms are given only by Eq.~\smssma,
the resulting 
QIRFP pattern has been discussed by previous authors. As mentioned earlier,
we can reproduce this case by setting 
$m_{\psi} = \Ataubar = \Abbar = \Atbar = \mu$ and
$m_{1,2}^2\rightarrow m_{1,2}^2+\mu^2$. However, for ease of presentation we
start by analysing the case $m_{\psi} = \Ataubar = \Abbar = \Atbar = 0$; but it 
is straightforward to check that our results are still valid when we 
include the supersymmetric $\mu$ term as above. 
The most robust prediction is easily seen to be that (at small $\tan\beta$)
\eqn\QIRsum{
\frakk{\Sigma (M_3^2)}{M_3(M_Z)^2}
\approx\delta\left[{(tf-F)^2\over{F^2}}
+{d\over{dt}}\left({t^2\over f}{df\over{dt}}\right)
-{t^2\over F}{df\over {dt}}\right]\biggr|_{M_Z},}
where 
\eqn\QIRdel{\delta = \left(\frakk{\alpha_0}{\alpha_3 (M_Z)}\right)^2}
and we have used 
$\sum_iC^t_if_i={d\over{dt}}\left({t^2\over f}{df\over{dt}}\right)$.
There is an analogous expression for large $\tan\beta$. So we see that
for large $y_0$, $\Sigma$ is independent of the initial values of the 
soft masses. 
The result 
\eqn\QIRmnum{
{\Sigma(M_Z)\over{M_3(M_Z)^2}}\approx 
\left\{\matrix{0.75&\hbox{small $\tan\beta$}\cr 0.76&\hbox{large $\tan\beta$}}
\right.}
(note the negligible difference between the large and small $\tan\beta$ 
cases) is in agreement with Refs.~\abel\kaza.

If we assume a universal scalar $(\hbox{mass})^2$, $m_0^2$,
at $M_U$ then  it is easy to see that there are similar fixed 
points for  the following quantities:

at small $\tan\beta$:
\eqna\QIRmore
$$\eqalignno{
\frakk{m_Q^2+m_2^2}{M_3(M_Z)^2} &\approx 0.28,&\QIRmore a \cr
\frakk{m_1^2+2m_2^2}{M_3(M_Z)^2} &\approx -0.75,&\QIRmore b\cr
\frakk{m_{\tbar}^2}{M_3(M_Z)^2} &\approx 0.47,&\QIRmore c\cr}$$
in broad agreement with Refs.~\abel,\kaza;

while at large $\tan\beta$: 
\eqna\QIRmorea$$\eqalignno{
\frakk{m_Q^2-m_{\tbar}^2}{M_3(M_Z)^2} &\approx 0.05,&\QIRmorea a\cr  
\frakk{2m_Q^2+m_2^2}{M_3(M_Z)^2} &\approx 0.81,&\QIRmorea b\cr}$$
which do not seem to appear explicitly in the literature, although
it is easy to see that, for example, they are implied by Eqs.~(20)-(23) 
of Ref.~\marcela.
Note also that, writing 
\eqn\QIRb{\eqalign{ \frakk{\Delta(M_Z)}{M_3^2
(M_Z)} &\approx  \delta\left[{(tf-F)^2\over{F^2}}-{t^2\over F}{df\over
{dt}}\right]\biggr|_{M_Z} -\delta\frakk{\Sigma (0)}{M_0^2}\cr &\approx -0.94 -
0.12\frakk{\Sigma (0)}{M_0^2}\cr}} 
then as long as $\frakk{\Sigma(0)}{M_0^2} < 7$  
then the dependence on $\Sigma (0)$ of this ratio is
suppressed.  The result is further QIRFP behaviour, for a {\it limited}
range  of boundary conditions at $M_U$ for the soft masses\kaz; we will
not discuss  this possibility further, however. 

As we pointed out before, the above predictions remain valid when the 
non-holomorphic terms simply reproduce the supersymmetric $\mu$-term. 
Let us turn now to examine the extent to which they 
survive the introduction of completely general non-holomorphic terms; 
firstly in the small $\tan\beta$ case. 
We see that 
$Y(t)$ in Eq.~\Ydef\ still depends on $m_{\psi}(0)$ and $\Atbar(0)$
as $y_0\rightarrow\infty$, and this dependence 
in fact grows with $y_0$, since the integrand of
$\Omega_{(6,\frak32)}$ develops a pole at $\tau=0$ as $y_0\rightarrow
\infty$; similarly for $m_1^2$. Clearly, however, 
since $\Sigma$ is independent of $Y$, the results 
Eqs.~\QIRsum\ and \QIRmnum\ survive in the general case, but not 
Eq.~\QIRmore{}. 

For large $\tan\beta$, we find that 
for $\tY$ in Eq.~\tYdef\ we have $\tY\approx \frak27m_{\psi}(0)^2$ 
as $y_0\rightarrow\infty$.  $\Sigma$, $m_Q^2-m_{\tbar}^2$ and 
$2m_Q^2+m_2^2$ are, however, independent 
of $\tY$ so we obtain
\eqn\QIRsigg{{\Sigma(M_Z)\over{M_3(M_Z)^2}}\approx.76}
for arbitrary initial scalar masses, and
\eqn\QIRmoreab{\eqalign{
\frakk{m_Q^2-m_{\tbar}^2}{M_3(M_Z)^2} &\approx 0.05,\cr  
\frakk{2m_Q^2+m_2^2}{M_3(M_Z)^2} &\approx 0.81\cr}}
for a universal scalar  $(\hbox{mass})^2$. 
The fact that
the latter QIRFPs are valid even for non-supersymmetric 
$m_{\psi}$, $\Ataubar$, $\Abbar$ and $\Atbar$ is rather remarkable. 
It is clear
from Eq.~\solsb{}\  that this happens because in the limit $y_0\to\infty$, 
$m_{\psi}, \overline{A}_{t,b,\tau}$  all approach zero.
In Ref.~\jjnss\ we argued that in the presence of the non-holomorphic 
soft terms it might be that there was no explicit \sic\ $\mu$-term,
and we explicitly demonstrated that  
there were regions of parameter space corresponding to an 
acceptable electroweak vacuum. Unfortunately this scenario 
cannot be implemented here, since using $m_1^2\approx m_2^2$ 
we obtain at once  (using the tree minimisation conditions 
in the absence of a $\mu$-term) that 
$m_1^2\approx m_2^2\approx -\frak{1}{2}M_Z^2$ which violates 
the well known requirement that $m_1^2 + m_2^2 > |m_3^2|$.

%
The new parameters themselves do exhibit QIRFP  behaviour 
if we consider  ratios of  $\overline{A}_{t,b,\tau}$ to $m_{\psi}$. 
Starting with the small $\tan\beta$ case, we see that  while
${\cal A}_t$, ${\cal A}_b$ and ${\cal A}_{\tau}$ 
have no individual QIRFP, we have (as $y_0\to\infty$)  
\eqn\sumrule{
{{\cal A}_t}+{{\cal A}_b}\approx 2.} 
\vfill
\eject
As
pointed out in Ref.~\jjnss\ and clearly manifested in Eqs.~\sols{}, the ratios 
of $\overline{A}_{t,b,\tau}$ to $m_{\psi}$ have true infra-red fixed
points  (i.e. as $t\rightarrow\infty$) of 1, corresponding to the
supersymmetric limit,  and so
${{\cal A}_t}+{{\cal A}_b}$ has an
infra-red fixed  point of 2. The point is that the QIRFP behaviour 
occurs for finite $t$ rather than for $t\rightarrow\infty$. In Fig.~1 we show
the approach to the QIRFP for ${{\cal A}_t}+{{\cal A}_b}$ 
for $\tan\beta$ close to the QIRFP value. There is clear convergence towards the
QIRFP, although this convergence is somewhat slowed by the power $\frak16$
of $H_7(t,y_0)$ in Eq.~\sols{c}. This means that to see significant convergence
we need to be at or beyond the limit of perturbative believability for $y_0$
(though in Ref.\jjpf\ we argued using Pad\'e-Borel summation techniques that 
the domain of attraction of the QIRFP could be extended beyond the na\"ive 
perturbative region).

In Fig.~2 we show 
the contrasting behaviour of the
individual ratio ${{\cal A}_b}$ which clearly has no QIRFP; the approach to the 
fixed point value ${\cal A}_b =1$ is much slower than the approach to the
QIRFP in Fig.~1.
\smallskip
\epsfysize= 2.5in
\centerline{\epsfbox{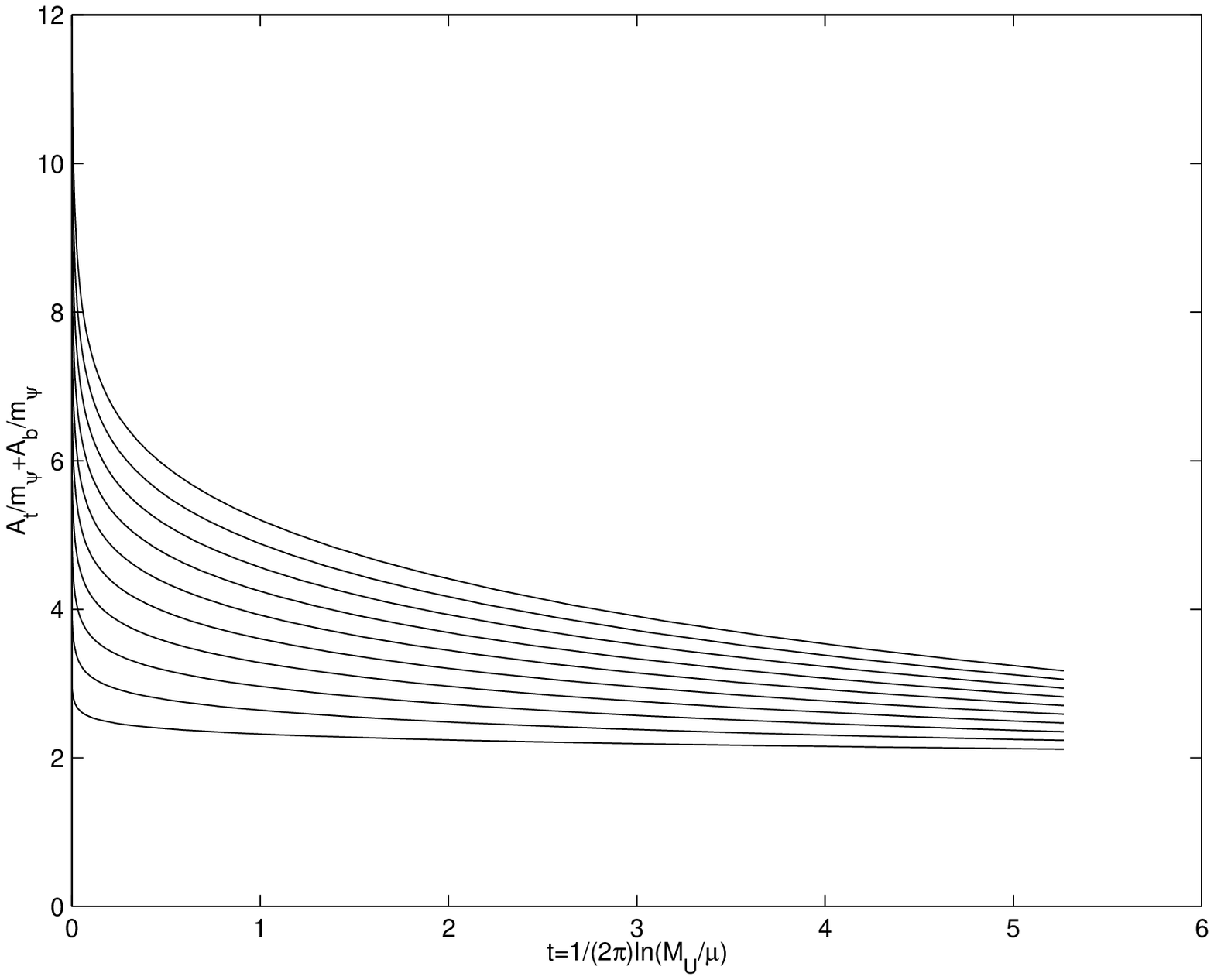}}
\in
{\it \noindent Fig.~1:}
A plot of ${{\cal A}_t}+{{\cal A}_b}$ against $t={1\over{2\pi}}
\ln\left({M_U\over{\mu}}\right)$ for $\tan\beta\approx1.7$, 
with
${\cal A}_{b}(M_U)={\cal A}_{\tau}(M_U)=1$, and with $2\le{\cal A}_t(M_U)
\le 11$.
\medskip
\out
\smallskip
\epsfysize= 2.5in
\centerline{\epsfbox{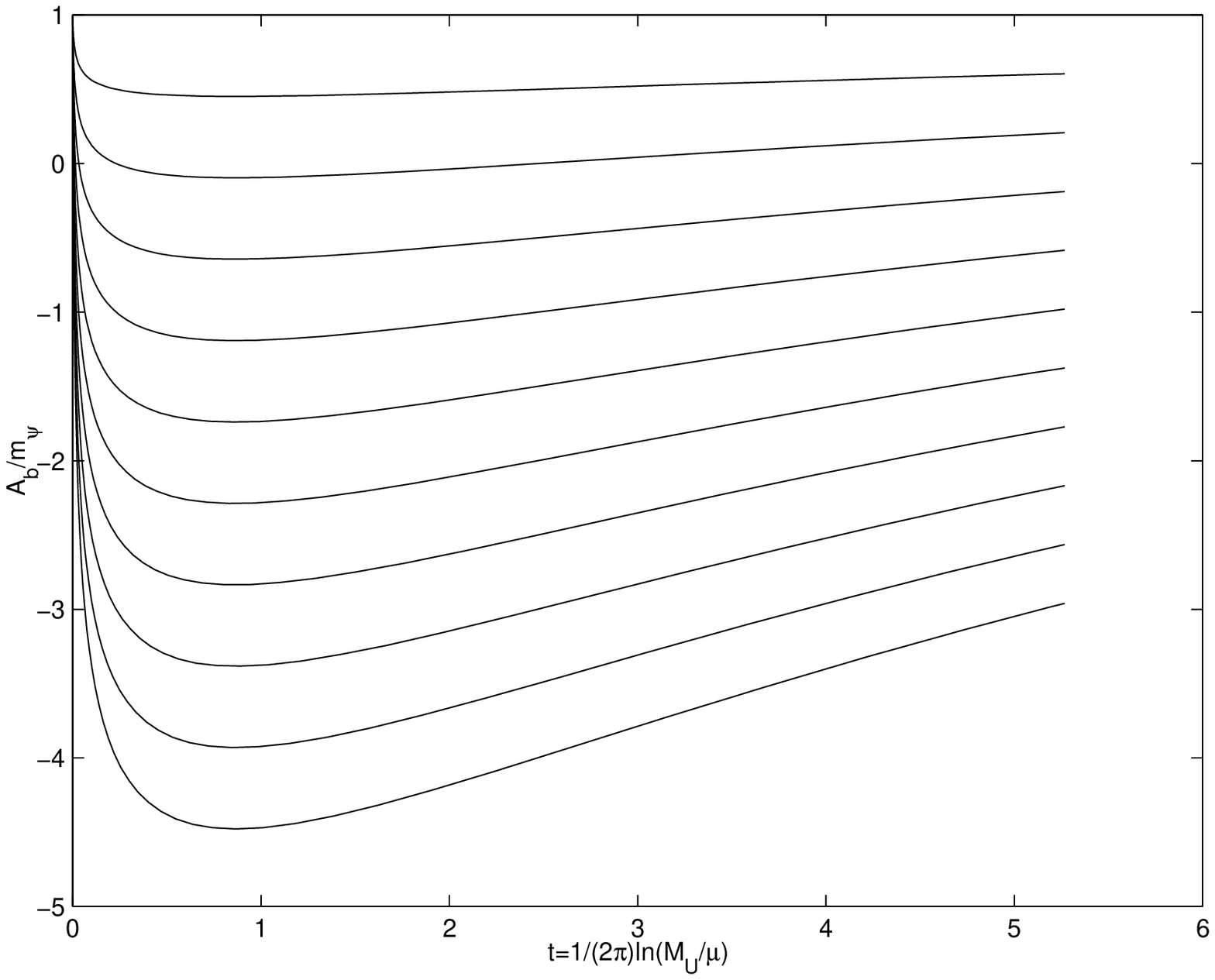}}
\in
{\it \noindent Fig.~2:}
A plot of ${{\cal A}_b}$ against $t={1\over{2\pi}}
\ln\left({M_U\over{\mu}}\right)$ for $\tan\beta\approx1.7$, with
${\cal A}_{b}(M_U)={\cal A}_{\tau}(M_U)=1$, and with $2\le{\cal A}_t(M_U)
\le 11$. 
\medskip
\out
Of course for the prediction Eq.~\sumrule\ to have experimental
relevance  we would need $m_{\psi}$ to be non-negligible at $M_Z$:
otherwise, the associated contributions to the squark mass matrices
would be small.  Since  as we already remarked, in fact $m_{\psi}(t)\to
0$ as $y_0\to\infty$, it  follows that we would need $m_{\psi}$ to be
large at $M_U$.  Therefore we cannot  simultaneously have good fixed point
convergence  for the $m^2/M_3^2$ fixed points and the ${\cal A}$ fixed
point,  Eq.~\sumrule, {\it and\/}  have the latter have experimental
consequences. An exception is  Eq.~\QIRmorea{a}, since {\it
both\/} $\tDelta$ and $\tY$ cancel in this  combination, as is easily
seen from Eq.~\msofta.

In the large $\tan\beta$ case, we see that
if we have ${{\cal A}_t}(0) = {{\cal A}_b}(0)$ then 
there is a QIRFP  ${{\cal A}_t} = {{\cal A}_b} = 1$, while
${{\cal A}_{\tau}}$ actually grows for large $y_0$, unless 
\eqn\eqatau{
5{{\cal A}_{\tau} (0)}+{{\cal A}_t (0)}
-4{{\cal A}_b (0)}= 2.}
This behaviour reflects the fact that the stability matrix for the evolution
of ${{\cal A}_t}$, ${{\cal A}_b}$ and ${{\cal A}_{\tau}}$, 
given in Ref.~\jjnss,  has at least one negative
eigenvalue in this case.  

\newsec{Summary}

In this paper we have continued the study of the RG evolution of 
``non-holomorphic'' soft terms that we began in Ref.~\jjnss.
In a special class of theories, we 
have shown the existence of a relation between 
the $r$-term and $m^{ia}$ term that is RG-invariant, 
at least through two loops. 

We have also explored the infra-red 
behaviour of these soft terms in the MSSM. Of course, in general 
we simply have a much enlarged parameter space, so we 
have restricted our attention to the two cases when either the 
top-quark Yukawa is close to its quasi-infra-red fixed point 
(corresponding to small $\tan\beta$) or when the top and 
bottom Yukawas are equal and close to a quasi-infra-red fixed 
point (corresponding to large $\tan\beta$)

We have shown that (for small $\tan\beta$) we obtain the predictions
at $M_Z$ (independent of the boundary conditions at $M_U$)
\eqna\conca$$\eqalignno{
m_Q^2 + m_{\tbar}^2 + m_2^2 &\approx.75M_3^2&\conca a\cr
A_t &\approx -0.6M_3 &\conca b\cr
\Atbar + \Abbar &\approx 2m_{\psi} &\conca c,\cr}$$
where Eq.~\conca{c}\ certainly holds but for Eqs.\conca{a,b}\ to 
hold it would have to be that $m_{\psi} << M_3$.  

For large $\tan\beta$ Eq.~\conca{}\ again holds (with the same qualification), 
but in addition we also have (if there is a universal $m_0^2$ at $M_U$) 
\eqn\concb{\eqalign{
{m_Q^2-m_{\tbar}^2}&\approx 0.05{M_3(M_Z)^2} \cr  
{2m_Q^2+m_2^2} &\approx 0.81{M_3(M_Z)^2}.\cr}}

Finally we note that 
recently an interesting phenomenon termed ``focussing'' has been 
noticed\ref\feng{J.L.~Feng and T.~Moroi, hep-ph/9907319\semi J.L.~Feng,
K.T.~Matchev and T.~Moroi, hep-ph/9909334}; this  also confers a
substantial measure of predictivity on the values of certain  soft
masses. In focussing, the value of some soft mass at a particular scale
is independent of the soft mass scale at unification. For a certain
class of boundary conditions at unification, which includes the usual
``universal''  case, this focus point of the RG trajectories occurs for
$m_2^2$ and at a value close to the weak scale (for a range of moderate
values of $\tan\beta$). We  note that, in contrast to the QIRFP case,
focussing is not driven by the behaviour of the Yukawa couplings at
unification. 
\bigskip\centerline{{\bf Acknowledgements}}\nobreak

This work was supported in part by a Research Fellowship from the 
Leverhulme Trust; and part of it was done at the UK Theory 
Institute at Swansea.  One of us (TJ) thanks Marcela Carena for 
a conversation, and in particular for drawing to our attention 
the treatment of the large $\tan\beta$ case from Ref.~\marcela. 

\appendix{A}{}

In this appendix we list the results for the two-loop $\beta$-functions for 
$r^{jk}_i$, $m_A^{ia}$, $b^{ij}$ and $(m^2)^i{}_j$, with $m_F$ set to zero.
(The two-loop
$\beta$-functions for $M$ and $h^{ijk}$ may be found in 
Refs.~\ref\jj{I.~Jack and
D.R.T.~Jones, \plb 333 (1994) 372}\ref\mv{S.P.~Martin
and M.T.~Vaughn, \prd50 (1994) 2282\semi
Y.~Yamada, \prd50 (1994) 3537\semi
I.~Jack, D.R.T.~Jones,
S.P.~Martin, M.T.~Vaughn and Y.~Yamada, \prd50 (1994) R5481}).) We find
\eqn\betar{\eqalign{
\llf (\beta^{(2)}_r)^{jk}_i &= -2g^2Y^{jpl}Y_{ipm}C(R)^m{}_n\rt^{kn}_l
-2g^2Y^{jpl}Y_{ipm}\rt^{km}_nC(R)^n{}_l
\cr
&-2Y^{jlm}Y_{iln}Y_{mpq}Y^{npr}\rt^{kq}_r-2g^4\rt^{jk}_m[C(R)^2]^m{}_i\cr
&-2g^4[C(R)^2]^j{}_l\rt^{kl}_i-4g^4\rt^{jl}_mC(R)^k{}_lC(R)^m{}_i
-2g^2\rt^{jl}_mY_{iln}Y^{mnq}C(R)^k{}_q\cr
&-2g^2\rt^{jl}_mY_{lnp}Y^{kmn}C(R)^p{}_i-Y^{jkl}Y_{lmn}Y^{nqr}Y_{ipr}\rt^{mp}_q
\cr
&-Y^{jlm}Y_{iln}Y^{knp}Y_{mqr}\rt^{qr}_p-2Y^{jpq}Y_{mnp}Y^{klm}Y_{ilr}\rt^{nr}_q
\cr
&-3g^2Y^{jkn}Y_{lmn}\rt^{lm}_pC(R)^p{}_i
-2g^2Y^{jlm}Y_{iln}\rt^{np}_mC(R)^k{}_p\cr
&+4g^2Y^{jlm}(R_a)^n{}_lY_{inp}\rt^{pq}_m(R_a)^k{}_q
+2g^2\rt^{lm}_nC(R)^j{}_mY^{knq}Y_{ilq}\cr
&-2g^2Y^{jkq}Y_{npq}\rt^{mp}_l(R_a)^n{}_i(R_a)^l{}_m
+4g^4\rt^{jm}_i[C(R)^2]^k{}_m\cr
&+8g^4\rt^{jl}_m(R_aR_b)^m{}_l(R_aR_b)^k{}_i
-4g^4C(G)\rt^{jl}_m(R_a)^m{}_l(R_a)^k{}_i\cr
&-g^2\left[\rt^{lm}_nY_{lmp}Y^{jpq}+2\rt^{lq}_mY_{lnp}Y^{jpm}-2g^2C(G)\rt^{jq}_n
\right](R_a)^n{}_q(R_a)^k{}_i\cr
&-4g^2(R_a)^l{}_m\rt^{jm}_nP^n{}_l(R_a)^k{}_i-\rt^{jl}_iY_{lmn}Y^{kmp}P^n{}_p
\cr
&-Y_{lmn}Y^{jkn}\rt^{mp}_iP^l{}_p-2g^2\rt^{jl}_i[C(R)P]^k{}_l
-g^2\rt^{jk}_l[C(R)P]^l{}_i\cr
&-2Y^{jlm}Y_{inp}P^n{}_l\rt^{kp}_m-2Y^{jlm}Y_{iln}\rt^{nk}_pP^p{}_m
-2Y^{jlm}Y_{iln}P^n{}_p\rt^{kp}_m\cr
&-g^4Q\rt^{jk}_lC(R)^l{}_i+2g^4Q\rt^{jl}_iC(R)^k{}_l
-\frak12\rt^{jk}_lY^{lmn}Y_{imp}P^p{}_n\cr
&+\sqrt2g\Bigl\{6g^4C(G)Q(R_a)^j{}_im_A^{ak}
-4g^2\tr[PR_aR_b](R_b)^j{}_im_A^{ak}\cr
&-2g^2C(G)(Pm_A^a)^j(R_a)^k{}_i-(R_a)^j{}_iY_{lmn}Y^{kmp}P^n{}_pm_A^{al}\Bigr\}
\quad+j\leftrightarrow k,\cr}}
\eqn\betam{\eqalign{
\llf (\beta^{(2)}_{m_A})^{ai} &= -2g^2\tr[PR_aR_b]m_A^{ib}
-Y^{ikl}Y_{kmn}P^m{}_lm_A^{na}-2\sqrt2gY^{ikl}Y_{kmn}(R_a)^m{}_p\rt^{np}_l\cr
&+2g^2C(G)[2g^2Qm_A^{ia}-(Pm_A^a)^i-\sqrt2g(R_a)^k{}_l\rt^{il}_k],\cr}}
where $P^i{}_j$ and $Q$ are as defined in Eq.~\Aac, and 
\eqn\rtdef{
\rt_i^{jk}  = r_i^{jk}-\sqrt{2}g\left[(R_a)^j{}_i m_A^{ka}
+ (R_a)^k{}_i m_A^{ja}\right].}
Clearly, on the RG trajectory given by Eq.~\rfin\ (now with $\mu=0$)
$(\beta^{(2)}_{m_A})^{ai}$ 
and especially $(\beta^{(2)}_r)^{jk}_i$ simplify considerably, and satisfy Eq.~\RGa.
We further find
\eqn\betab{\eqalign{
\llf(\beta_b^{(2)})^{ij} &= 
-b^{il}Y_{lmn}Y^{mpj}P^n_p-2g^2C(R)^i{}_kV^{kj}
+2g^4b^{ik}C(R)^j{}_kQ\cr&
+8g^4C(G)[T(R)-2C(G)]m_A^{ia}m_A^{ja}
-4g^2(R_bR_a)^i{}_kY^{jkl}Y_{lmn}m_A^{ma}m_A^{nb}\cr&
+2g^2C(G)Y^{ijk}Y_{klm}m_A^{la}m_A^{ma}
-2Y^{ikn}Y_{klm}r^l_{np}h^{mpj}
-Y_{kln}Y^{inp}r_p^{mj}r^{kl}_m\cr&
-2Y^{imn}Y_{klm}r^{lp}_nr^{jk}_p
-2r^{ik}_lr^{jl}_mP^m{}_k
+2g^2[C(R)^l{}_mr^{ik}_l-C(R)^k{}_lr^{il}_m]r^{jm}_k
\cr&
-Y^{ikl}X^m{}_lr^j_{km}
-2h^{ikl}P^m{}_lr^j_{km}\quad+j\leftrightarrow k,\cr}}
\eqn\betamsoft{\eqalign{
\llf(\beta^{(2)}_{m^2})^i{}_j &= \biggl(-\Bigl[(m^2)_j{}^lY_{lmn}Y^{mpi}
+\frak12Y_{jlm}Y^{ipm}(m^2)^l{}_n+\frak12Y_{jnm}Y^{ilm}(m^2)^p{}_l
\cr&+Y_{jln}Y^{irp}(m^2)^l{}_r+h_{jln}h^{ilp}\cr&
+4g^2MM^*C(R)^i{}_n\delta^p{}_j+2g^2(R_a)^i{}_j(R_am^2)^p{}_n\Bigr]P^n{}_p\cr
&+\bigl[2g^2M^*C(R)^p{}_j\delta^i{}_n-h_{jln}Y^{ilp}\bigr]X^n{}_p \cr
&-\frak12\bigl[Y_{jln}Y^{ilp}+2g^2C(R)^p{}_j\delta^i{}_n\bigr]W^n{}_p
+12g^4MM^* C(R)^i{}_jQ\cr
&+4g^4SC(R)^i{}_j
+2Y^{ikl}Y_{jmn}[P^m{}_k+g^2C(R)^m{}_k]m_{Aak}m_A^{an}
\cr
&+4g^2(R_bR_a)^i{}_j
m_{Aa}Pm_A^b-4g^2(R_bR_aP)^i{}_jm_{Aa}m_A^b
\cr
&+4Y^{ikl}Y_{jkm}(m_A^aP)_lm_A^{am}-2g^2C(G)Y^{ikl}Y_{jkm}m_{Aal}m_A^{am}
\cr
&+8g^4Q(R_bR_a)^i{}_jm_{Aka}m_A^{kb}
+4g^4C(G)(R_bR_a)^i{}_jm_{Aka}m_A^{kb}\cr&
+4g^4C(G)[2Q+3C(G)]m_A^{ia}m_{Aja}
+2g^2Y^{ikl}C(R)^m{}_jY_{kmn}m_{Ala}m_A^{na}\cr&
-8g^4(R_aR_b)^i{}_jm_{Ac}R_aR_bm_A^c
+2Qg^2Y^{ikl}Y_{jkm}m_{Ala}m_A^{am}\cr&
+8g^2(R_aR_b)^i{}_kY_{jlm}Y^{kln}m_{Anb}m_A^{ma}\cr&
+8g^2(R_a)^i{}_k Y_{jlm}\left(Y^{kln}(m_{Ab}R_a)_nm_A^{bm}
+(R_a)^l{}_nY^{nkp}m_{Apb}m_A^{mb}\right)\cr
&+4g^2(R_a)^i{}_k(R_a)^l{}_jY_{lmn}Y^{kmp}m_{Apb}m_A^{nb}
\cr
&+4\sqrt2\left(gr^{ik}_lY_{jmn}(R_a)^n{}_kY^{lmp}m_{Apa}
+2g^3(R_aR_b)^i{}_kr^{kl}_j\left(m_{Ab}R_a\right)_l\right)\cr
&-2Y^{ikl}Y_{lnp}r^{mn}_jr^p_{km}
-Y_{jkm}Y^{mpq}r^{ik}_lr^l_{pq}\cr
&-2Y^{ikl}Y_{kmp}r^{pn}_lr^m_{jn}-4g^2C(R)^i{}_mr^{kl}_jr^m_{kl}\cr
&-2g^2[C(R)^m{}_lr^{ik}_m-C(R)^k{}_mr^{im}_l]r^l_{kj}
-4g^2(R_a)^i{}_k(R_a)^m{}_nr^{kl}_jr^n_{ml}\cr
&-r^{kl}_jr^i_{km}P^m{}_l-r^{ik}_lr^l_{jm}P^m{}_k-r^{ik}_lr^m_{jk}P^l{}_m
\biggr)+{\rm h.c.},\cr}}
where $V^{ij}$, $W^i{}_j$ and $X^i{}_j$  
are as defined in Eqs.~\Aoldm{b,c}, \Aab\ but with $m_F=0$, and
where 
\eqn\Ava{
S\delta_{ab}=(m^2)^i{}_j(R_aR_b)^i{}_j-MM^*C(G)\delta_{ab}.}
The form of Eqs.~\betab\ and \betamsoft\ on the RG-trajectory is less 
clear than in the case of Eqs.~\betar\ and \betam, but nevertheless after 
some work we find that Eqs.~\RGc\ and \RGe\ are satisfied at this order
(with $\mu=0$). (In
the case of Eq.~\betamsoft, we are again obliged to specialise to theories
for which the matter multiplet
satisfies $C(R)^i{}_j = C(G) \delta^i{}_j$, as at one loop.) 
{\it A fortiori}, we see that the same conditions which imply
one-loop finiteness also guarantee two-loop finiteness, as was discovered in
the case of the standard soft couplings in Ref.~\jj. Although we have 
presented two-loop results for the case $m_F=0$, we have checked that for 
$m_F=\mu$, the relations Eqs.~\rfin, \RGb\ and \RGd\
continue to be RG-invariant--in other words, the relations Eq.~\RGee{} are 
RG-invariant in a theory with a 
supersymmetric $\mu$-term together with $L_{\rm SOFT}$ as in Eq.~\Lsoft, 
and with $m_F = 0$ in Eq.~\Aaf.  As explained earlier, this is a 
consequence of the fact that
couplings satisfying Eqs.~\RGee{}\ follow from the single holomorphic term
Eq.~\spuri.
\listrefs
\bye